\documentclass{INTERSPEECH2023}
\usepackage{multirow}
\usepackage{subcaption}
\usepackage{hhline}

\interspeechcameraready 

\title{RAMP: Retrieval-Augmented MOS Prediction via Confidence-based Dynamic Weighting}

\name{Hui Wang$^1$, Shiwan Zhao$^\dag$, Xiguang Zheng$^2$,Yong Qin$^{1,*}$}
\address{
  $^1$Nankai University, China\\
  $^2$Kuaishou Technology, China 
  }
\email{wanghui\_hlt@mail.nankai.edu.cn, zhaosw@gmail.com, zhengxiguang@kuaishou.com, qinyong@nankai.edu.cn}

\begin{document}

\maketitle

\renewcommand{\thefootnote}{}
\footnotetext{\dag Independent researcher.}
\footnotetext{* Corresponding author.}

\begin{abstract}
Automatic Mean Opinion Score (MOS) prediction is crucial to evaluate the perceptual quality of the synthetic speech. While recent approaches using pre-trained self-supervised learning (SSL) models have shown promising results, they only partly address the data scarcity issue for the feature extractor. This leaves the data scarcity issue for the decoder unresolved and leading to suboptimal performance. To address this challenge, we propose a retrieval-augmented MOS prediction method, dubbed {\bf RAMP}, to enhance the decoder's ability against the data scarcity issue. A fusing network is also proposed to dynamically adjust the retrieval scope for each instance and the fusion weights based on the predictive confidence. Experimental results show that our proposed method outperforms the existing methods in multiple scenarios.
\end{abstract}

\section{Introduction}
\label{sec:intro}

Evaluation of synthetic speech typically involves the use of objective and subjective methods. Objective methods, such as Mel Cepstral Distortion (MCD) and F0 Frame Error (FFE), require reference audio, making them impractical or even impossible to use in scenarios where reference audio is unavailable \cite{Chinen2022UsingRA}. On the other hand, subjective methods, such as Mean Opinion Score (MOS), rely on listening tests conducted by crowdsourced listeners, which can be time and resource consuming.

Automatic MOS prediction using machine learning \cite{lo2019mosnet, Patton2016AutoMOSLA, Fu2018QualityNetAE} has recently gained popularity as it allows for quality evaluation that matches human perception with low time and resource cost. However, training of such methods still relies on manually scored data and often suffers from data scarcity due to limited budgets. To alleviate this problem, self-supervised learning (SSL) models \cite{baevski2020wav2vec, Hsu2021HuBERTSS} trained on large-scale unlabeled data are employed as the feature extractor, followed by a downstream MOS prediction decoder trained on small-scale data with labeled MOS scores \cite{Cooper2021GeneralizationAO, Ragano2022ACO}. Recent works based on this SSL-based structure \cite{Tseng2022DDOSAM, Saeki2022UTMOSUS, Yang2022FusionOS} have outperformed the earlier works trained from scratch \cite{lo2019mosnet,leng2021mbnet,huang2022ldnet} in the MOS prediction tasks.

\begin{figure}[t]
  \centering
  \includegraphics[width=\linewidth]{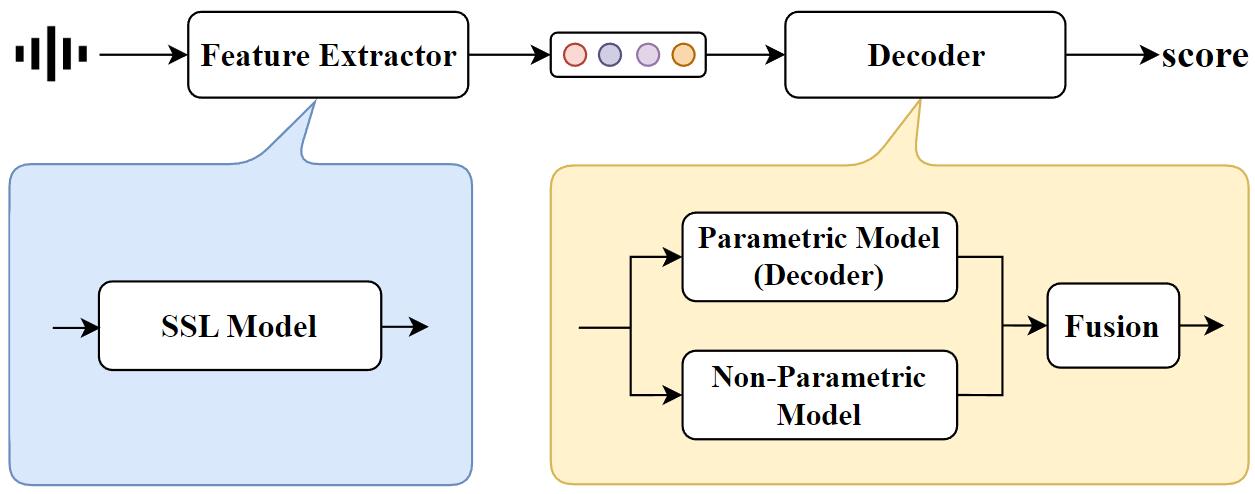}
  \caption{
  The SSL-based framework (bottom left) improves the feature extractor with an SSL model and the proposed framework (bottom right) augments the decoder by introducing a non-parametric model and a fusing network.}
  \label{fig:framework}
  \vspace{-10pt}
\end{figure}

Existing SSL-based frameworks, as depicted in Figure~\ref{fig:framework},  
mainly focus on improving the feature extractor based on the help of large-scale pre-training corpora to obtain better representations. To name a few, Yang et al. \cite{Yang2022FusionOS} fuse seven pre-trained SSL models, as both the data and model configurations used for pre-training can impact SSL model performance. Tseng et al. \cite{Tseng2022DDOSAM} demonstrate that domain adaptive pre-training (DAPT) can reduce domain mismatch between speech in the pre-training corpus and fine-tuning MOS corpus. Saeki et al. \cite{Saeki2022UTMOSUS} concatenate as much information as possible from phonemes, raters, and other sources with the SSL representation vector. Vioni et al. \cite{Vioni2022InvestigatingCN} include prosodic and linguistic features as inputs to improve system performance. 

However, the decoder, which decodes the features into scores, is only trained with the MOS dataset. The data scarcity issue for the decoder remains unsolved, leading to suboptimal performance. A few works attempt to improve the neural decoder. Tseng et al. \cite{Tseng2022DDOSAM} replace the linear layer of the decoder with a DNN, but their results indicate that increasing the decoder's parameters does not necessarily improve performance. Chen et al. \cite{Chen2021InQSSAS} design multi-task heads to predict both quality and intelligibility scores simultaneously. 
In summary, these neural decoders can be classified as parametric methods by learning the mapping from representations to scores, which generally require large-scale labeled data to achieve good generalization ability and tend to suffer difficulties in adapting to new domains with distribution shifts.

\begin{figure*}[t]
  \centering
  \includegraphics[width=\linewidth]{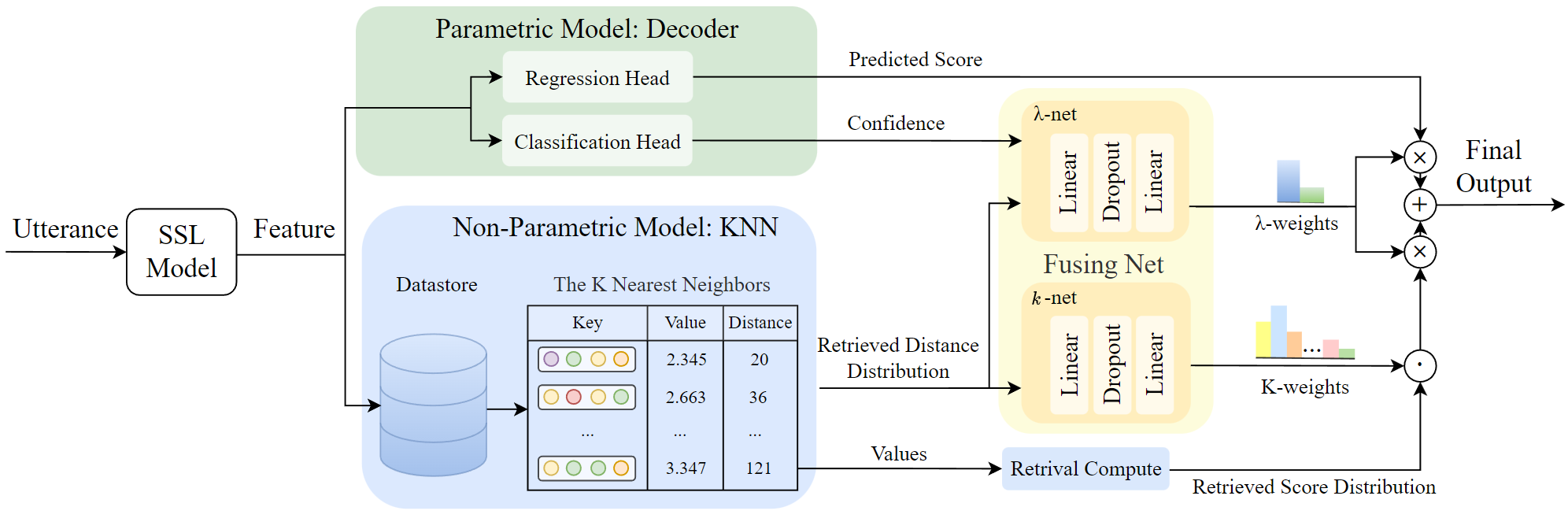}
  \caption{The illustration of the system flow at the inference time.}
  \label{fig:system_flow}
  \vspace{-12pt}
\end{figure*}

Given the powerful SSL-based feature extractor and the weak decoder, we share a similar hypothesis with $k$NN-LM \cite{Khandelwal2019GeneralizationTM} that \emph{the representation learning is easier than the prediction}. Thus, we propose a retrieval-augmented MOS prediction method, dubbed {\bf RAMP}, to enhance the decoder for addressing the data scarcity issue. Specifically, as shown in Figure~\ref{fig:framework}, we augment the decoder of the SSL-based framework by linearly interpolating its prediction score with the $k$-nearest neighbors ($k$NN) model. 
The $k$NN model, which is a non-parametric method, not only is good at memorizing rare patterns \cite{Wan2022RescueIA} but also handles cross-domain issues in a flexible way \cite{Khandelwal2019GeneralizationTM}.

The fusion of the parametric and non-parametric methods has been successfully utilized in tasks such as language modeling \cite{Khandelwal2019GeneralizationTM, Guu2020REALMRL} and machine translation \cite{Khandelwal2020NearestNM}. However, they combine the two methods by static weights. We argue that the predictive ability of parametric methods varies for different instances due to the uneven distribution of data in the dataset. For instance, the model may be more confident in predicting scores for some instances than for others. In such cases, less help from non-parametric methods may be needed. To this end, we propose a \emph{confidence-based dynamic weighting} scheme to balance the outcomes of the two methods. Furthermore, the number of retrievals of the $k$NN model is preset and sensitive to noise \cite{Zheng2021AdaptiveNN, Jiang2022TowardsRK}. To improve the robustness of the model, we automatically predict the importance of the number of retrievals and average the scores for different numbers. 

The main contributions of this work are threefold:
\begin{itemize}
\item We propose RAMP, a novel retrieval-augmented MOS prediction method, to enhance the neural decoder in SSL-based frameworks.
\item We design a fusing network to dynamically adjust the retrieval scope for each instance and the fusion weights based on the predictive confidence.
\item We have demonstrated the effectiveness of the method through extensive comparative and ablation experiments.
 \end{itemize}


\vspace{-6pt}
\section{The System Overview}

The training process of our system consists of two stages. In the first training stage, we fine-tune the SSL-based model (i.e., the feature extractor) with multi-task heads (i.e., the decoder). In the second stage, we freeze the feature extractor and decoder to train the fusing network to combine results from the decoder and the $k$NN model. Note that we use the frozen feature extractor on training data to construct the datastore.

The overall flow of our system at the inference time is presented in Figure~\ref{fig:system_flow}. During inference, the SSL model extracts the feature representation for the input utterance. The features are fed into the parametric and non-parametric paths separately. The corresponding outputs are then fused to get the final output. In the case of evaluating cross-domain audios, the steps are identical, except replacing the datastore with the target domain data without an additional fine-tuning stage.

 
\begin{table*}[th]
    \vspace{-12pt}
    \caption{The performances of our framework along with two systems on BVCC and BC2019 test set.}
    \centering
    \label{tab:comparative}

    \setlength{\belowcaptionskip}{0pt}%
    \vspace{-6pt}
    \subcaption{The results of the in-domain experiments.} \label{tab:in-domain}
    \begin{tabular}{ c c c c c c c c c c }
    \toprule
        dataset& model& U\_MSE↓ & U\_LCC↑ & U\_SRCC↑ & U\_KTAU↑ & S\_MSE↓ & S\_LCC↑ & S\_SRCC↑ & S\_KTAU↑ \\ 
    \midrule
        \multirow{3}{*}{BVCC} & SSL-MOS & 0.246 & 0.875 & 0.872 & 0.697 & 0.113 & 0.928 & 0.923 & 0.770  \\ 
        & DDOS & 0.212 & 0.880  & 0.880  & 0.707 & 0.110  & \textbf{0.933} & \textbf{0.932} & 0.782 \\ 
         & RAMP  & \textbf{0.195}  & \textbf{0.881} & \textbf{0.881} & \textbf{0.708} & \textbf{0.097} & 0.931 & \textbf{0.932} & \textbf{0.784} \\
    \midrule
        \multirow{3}{*}{BC2019} & SSL-MOS & 0.253 & 0.901 & 0.871 & 0.690 & 0.098 & 0.980 & 0.970 & 0.871  \\ 
        & DDOS & \textbf{0.169} & 0.914  & 0.887  & 0.710 & \textbf{0.052}  & 0.976 & 0.955 & 0.848 \\ 
         & RAMP  & 0.188  & \textbf{0.916} & \textbf{0.891} & \textbf{0.717} & 0.053 & \textbf{0.987} & \textbf{0.987} & \textbf{0.926} \\
    \bottomrule    
    \end{tabular}

    \vspace{3mm}
    \subcaption{The results of the cross-domain experiments.} \label{tab:cross-domain}
    \begin{tabular}{ c c c c c c c c c c }
    \toprule
        dataset& model& U\_MSE↓ & U\_LCC↑ & U\_SRCC↑ & U\_KTAU↑ & S\_MSE↓ & S\_LCC↑ & S\_SRCC↑ & S\_KTAU↑ \\ 
    \midrule
    \multirow{4}{*}{BC2019} & SSL-MOS & 3.187 & 0.527 & 0.549 & 0.403 & 2.976 & 0.590 & 0.655 & 0.569  \\
    & DDOS & 1.331 & 0.678 & 0.694 & 0.502 & 1.119 & 0.766 & 0.797 & 0.637  \\
    & RAMP  & 0.658 & 0.826 & 0.780 & 0.587 & 0.493 & 0.929 & 0.907 & 0.772  \\
    & RAMP(np)  & \textbf{0.294} & \textbf{0.842} & \textbf{0.789} & \textbf{0.596} & \textbf{0.093} & \textbf{0.955} & \textbf{0.926} & \textbf{0.797}  \\
    \bottomrule    
    \vspace{-6pt}
    \end{tabular}
\vspace{-9pt}
\end{table*}

\vspace{-3pt}
\section{The Proposed Method}

 The proposed method consists of the parametric path, the non-parametric path, and the fusing network.

\vspace{-3pt}
\subsection{Parametric path}
The \emph{parametric path} refers to the use of a neural network-based decoder to handle representations. The decoder is designed as a multi-task architecture consisting of a regression head and a classification head, which are implemented using several linear layers following SSL-MOS \cite{Cooper2021GeneralizationAO}. The purpose of introducing a classification head is to guide the model during fine-tuning, as well as to output the confidence of each score bin. 
Obtaining the confidence of the parametric model is crucial for subsequent result fusion. For the $i$-th instance $(u_i, s_i)$, we first map the score $s_i$ to the bin id $b_i$. Then the model outputs a prediction score $S_p$ along with a confidence probability distribution $[c_1, c_2, ..., c_n]$ on $n$ bins. The loss function is defined by:
 %
 %
 \begin{equation} \label{eq:classficationloss}
    \mathcal{L}  = \mathcal{L}_{reg}(u_i, s_{i}) + \alpha \mathcal{L}_{cls}(u_i, b_{i}),
\end{equation}
where $\mathcal{L}_{reg}$ and $\mathcal{L}_{cls}$ are the MSE and cross entropy losses, for the regression and classification heads, respectively. 
The $\alpha$ is a hyper-parameter that balances the two types of losses.

\vspace{-3pt}
\subsection{Non-parametric path}
\label{sec:np}

Compared to the parametric path, the \emph{non-parametric path} directly exploits the representations with a $k$NN model. The representation of the utterance being evaluated is used as the query to retrieve the most similar data instances from the datastore, along with their corresponding distances and labels. This information is then used as part of the input to the fusing network.

\textbf{Datastore:} Let $f(\cdot)$ be the SSL model that maps an utterance $u$ to its representation. For a training sample ${(u_i,s_i)}\in D$, we create a key-value pair $(k_{i},v_{i})$, where the key $k_{i}=f(u_i)$ is the representation vector of the utterance $u_i$ and the value $v_i=s_i$ is its target score. The set of all key-value pairs constructed from all training examples in $D$ forms the datastore $(\mathcal K,\mathcal V)$:
\begin{align}
(\mathcal K,\mathcal V)=\{(f(u_i),s_i)|(u_i,s_i)\in D\}.
\end{align}

\textbf{Inference:}
During testing, the SSL model generates a representation $q$ for the input utterance, which is then utilized by the $k$NN model to retrieve the $k$ nearest neighbors, denoted by $\mathcal N_k$, from the datastore based on a given distance function $d(q,\cdot)$. The retrieved score $S_{r,k}$ is then computed as follows: 
\begin{align}
S_{r,k}=\sum_{(k_i, v_i) \in \mathcal{N}_k}w_{ik}v_{i},
\end{align}
where the weight $w_{ik}$ is the inverse of its distance. 
Thus, neighbors that are closer will exert more influence than those that are farther away.

Since $k$NN is sensitive to $k$, we improve its robustness by using various values of $k$ in the range $[1, 2, \ldots, K]$, where $K$ is the hyperparameter. As a result, we obtain the retrieved score distribution $[S_{r,1}, S_{r,2}, \ldots, S_{r,K}]$, as well as the retrieved distance distribution $[d_{1},d_{2},\ldots,d_{K}]$, where $d_{k}$ is the distance between $q$ and the $k$-th nearest neighbor.

\subsection{Fusing network}
\label{sec:fuse}

We introduce two lightweight networks, $k$-net and $\lambda$-net, to dynamically predict the probability of the retrieval scope $k$ and the fusion weight distribution $\lambda$ based on the predictive confidence for each instance. 

\textbf{\boldmath$k$-net:} The $k$-net is a lightweight net consisting of only two linear layers which dynamically predicts the probability of each $k$ for each instance. It takes as input the distance distribution $d=[d_{1}, d_{2}, \ldots, d_{K}]$ of the retrieved neighbors and outputs the probability distribution $p_{knet}$ over different values of $k$:
\begin{align}
p_{knet}=softmax({knet}(d)).
\end{align}
Therefore, the final retrieved score can be computed as a weighted average of different $S_{r,k}$:
\begin{align}
S_r=\sum_{k=1}^{K}p_{knet}(k) S_{r,k}.
\end{align}

\textbf{\boldmath$\lambda$-net:} 
The $\lambda$-net has the same structure as $k$-net. To balance the outcomes of two paths, it takes the confidences as the input, in addition to the distance distribution $d$. The confidences include the top-$8$ values $c^{top}$ from the confidence probability distribution $[c_1, c_2, ..., c_n]$, and two confidences $c^{S_r}$ and $c^{S_p}$ of the bins to which $S_r$ and $S_p$ belong. Thus, the weight and then the final score can be computed as:
%
\begin{align}
w=(w_{p},w_{r})&=softmax({\lambda net}([d;c^{top};c^{S_r};c^{S_p}])),\\
S &= w_{p}S_p + w_{r}S_r,
\end{align}
%
where, $S_p$ and  $S_r$ are the scores of the parametric and non-parametric paths, respectively.

\begin{table*}[]
\vspace{-6pt}
\caption{The performance of two paths in three datasets: SOMOS (\underline{b}ig), BVCC (\underline{m}edium), and BC2019 (\underline{s}mall).}
\label{tab:ablation1}
\tabcolsep=4pt
\resizebox{\linewidth}{!}{
\begin{tabular}{ccccccccccccccccc}
\toprule
\multicolumn{1}{l}{} & \multicolumn{2}{c}{U\_MSE↓} & \multicolumn{2}{c}{U\_LCC↑} & \multicolumn{2}{c}{U\_SRCC↑} & \multicolumn{2}{c}{U\_KTAU↑} & \multicolumn{2}{c}{S\_MSE↓} & \multicolumn{2}{c}{S\_LCC↑} & \multicolumn{2}{c}{S\_SRCC↑} & \multicolumn{2}{c}{S\_KTAU↑}\\ 

\multicolumn{1}{l}{} & P & NP & P & NP & P & NP & P & NP & P & NP & P & NP & P & NP & P & NP \\
\midrule
b & 0.185 & \textbf{0.179} & 0.658 & \textbf{0.668} & 0.653 & \textbf{0.658} & 0.468 & \textbf{0.473} & 0.182 & \textbf{0.174} & 0.664 & \textbf{0.676} & 0.659 & \textbf{0.665} & 0.474 & \textbf{0.480} \\
m & \textbf{0.200} & 0.201 & 0.874 & \textbf{0.880} & 0.876 & \textbf{0.880} & 0.702 & \textbf{0.708} & 0.106 & \textbf{0.101} & 0.917 & \textbf{0.934} & 0.915 & \textbf{0.934} & 0.756 & \textbf{0.787} \\
s & 0.226 & \textbf{0.196} & 0.880 & \textbf{0.904} & 0.831 & \textbf{0.868} & 0.634 & \textbf{0.682} & 0.064 & \textbf{0.045} & 0.968 & \textbf{0.984} & 0.926 & \textbf{0.972} & 0.797 & \textbf{0.889} \\
\toprule
\end{tabular}
}
\vspace{-6pt}
\end{table*}

\vspace{-3pt}
\section{Experiments}

\subsection{Datasets and Metrics}
The experiments in this paper use three corpora: BVCC \cite{cooper21_ssw}, BC2019 \cite{Wu2019TheBC}, and SOMOS \cite{Maniati2022SOMOSTS}. BVCC contains 7,106 English samples from the Blizzard Challenges, the Voice Conversion Challenges, and published samples of ESPNet \cite{Hayashi2020EspnetTTSUR}. The ratio of training/development/test is 70\%/15\%/15\%, respectively. BC2019 contains Mandarin TTS samples submitted to the 2019 Blizzard Challenge \cite{Wu2019TheBC}. There are 136 samples for training, 136 samples for validation, and 540 samples for testing. We also use SOMOS in the ablation study. It consists of 20K TTS audio files generated from several Tacotron-like acoustic models \cite{Wang2017TacotronTE} and an LPCNet vocoder \cite{Valin2018LPCNETIN}.

Model evaluation is performed at the utterance level and system level, denoted as `U\_' and `S\_', respectively. Mean square error (MSE) and various correlation coefficient metrics are used. In particular, the Linear Correlation Coefficient (LCC), the Spearman Rank Correlation Coefficient (SRCC) and the Kendall Tau Rank Correlation (KTAU) scores are evaluated. In general, smaller errors and higher correlations indicate better model performance.

\vspace{-3pt}
\subsection{Implementation Details}

We use the published wav2vec2.0 \cite{baevski2020wav2vec} base model pre-trained on Librispeech \cite{Panayotov2015LibrispeechAA} as the feature extractor. In both training stages, the models are trained for 1,000 epochs with a batch size of 4 and a learning rate of 0.0001. Training will be stopped early when the loss does not decrease for 20 epochs. Gradient accumulation is used to simulate large batch. In the experiments, the length of the score bin is 0.25. The $\alpha$ is set to 1 in equation \ref{eq:classficationloss}. We set $K=60$ for BVCC, $K=8$ for BC2019 in Table \ref{tab:comparative}. The L2 distance is used in Section \ref{sec:np}. 
We crop the long audio in the SOMOS dataset due to memory limitations for all compared methods.

\vspace{-3pt}
\subsection{In-domain Analysis}
We first compare the performance of our method RAMP with SSL-MOS \cite{Cooper2021GeneralizationAO} and DDOS \cite{Tseng2022DDOSAM}. SSL-MOS is one of the first systems to employ the SSL model, which delivers high prediction performance with an easy-to-use framework. DDOS is one of the winning solutions of VoiceMOS Challenge 2022 \cite{huang22f_interspeech}, providing competitive outcomes across many metrics.

Table \ref{tab:in-domain} shows the performance of three systems in BVCC and BC2019. On the BVCC test set, our system RAMP performs better than SSL-MOS in all metrics. Compared to DDOS, RAMP performs better in error metrics and similarly in correlation metrics. To evaluate the performance in BC2019, all three models are first trained in BVCC and fine-tuned in BC2019 following the same procedure of the VoiceMOS Challenge 2022 to make a fair comparison. 
On the BC2019 test set, RAMP also outperforms SSL-MOS in all metrics. While DDOS outperforms our system slightly in terms of error metrics, our system consistently performs well across all six correlation metrics.

\begin{figure}[t]
  \centering
  \includegraphics[width=\linewidth]{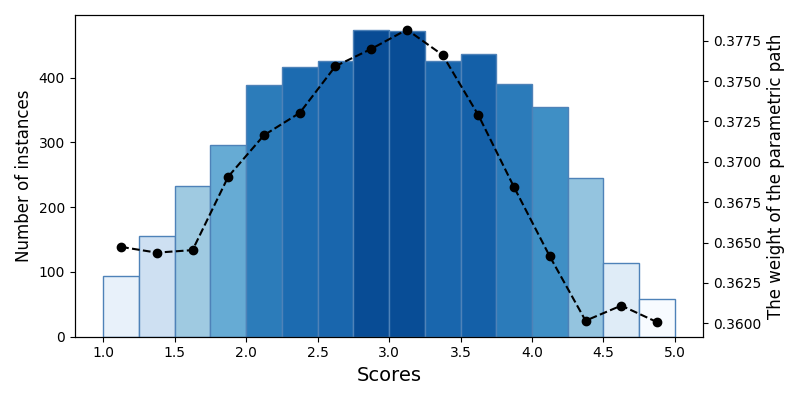}
  \vspace{-20pt}
  \caption{The bars are the score distribution of the BVCC dataset, and the curve represents the weights of the parametric path learned by the fusion network for each score bin. This shows dynamic weights are assigned aligning with the data distribution, indicating the effectiveness of the fusing net.}
  \label{fig:lambda}
  \vspace{-10pt}
\end{figure}

Moreover, in order to investigate how the confidence-based dynamic weighting scheme combines the two paths for efficient prediction, we visualize the fusion weights for various instances. Figure~\ref{fig:lambda} displays the score distribution of the BVCC data and the weights of the parametric path predicted by the fusing network. The data exhibit a long-tailed distribution with relatively small numbers of scores that are very low or very high. As mentioned in Section \ref{sec:intro}, this unbalanced distribution of data can cause the neural network to have different prediction capabilities for data located in different score intervals. For head data, the neural network-based decoder (i.e., the parametric path) can predict their scores with high confidence, requiring less help from the non-parametric path. Thus, smaller weights are assigned to the non-parametric path and higher weights to the parametric path. Conversely, for data at the tail, higher weights are assigned to the non-parametric path and smaller weights to the parametric path. 

\vspace{-3pt}
\subsection{Cross-domain Analysis}
The results of cross-domain experiments are presented in Table \ref{tab:cross-domain}. All systems are trained in BVCC and tested in the BC2019 test set without additional fine-tuning. As the results show, both SSL-MOS and DDOS exhibit a sharp drop in performance across domains. While our system can solve the cross-domain adaptation problem flexibly and excel in performance by updating the datastore without any additional training. 

To further investigate the help from the non-parametric path for cross-domain settings. We create a new variant called RAMP(np), which only includes the non-parametric path. This results in further improvements in performance, demonstrating that the distribution shift undermines the mapping from representations to scores, while the $k$NN model still performs well by directly leveraging the representations.


\subsection{Ablation study}


We first compare the performance of the parametric and non-parametric paths across different data scales. In Table \ref{tab:ablation1}, `P' and `NP' denote the parametric and non-parametric paths, respectively. We conduct the experiments on three datasets: SOMOS, BVCC, and BC2019, which correspond to tens of thousands, thousands, and hundreds of scales, respectively. We can observe that as the size of the dataset decreases, the non-parametric path boosts the performance more significantly. This demonstrates that our proposed method is more applicable to low-resource tasks like most MOS prediction tasks.

We then conduct an ablation study on the fusing net. Vanilla $k$NN uses a fixed $k$ for all instances and its performance is sensitive to the value of $k$. While our proposed fusing net specifies a predefined upper bound $K$ of the retrieval range. 
Then for each instance, it dynamically computes the weighted average score from the $K$ $k$NN models, each trained with different numbers of nearest neighbors from $1$ to $K$. The weight of each $k$NN is obtained through the network. We only show MSE and KTAU performance on the utterance level since the trends of system-level are similar. The results show that the fusing net performs better on average and performs more consistently when changing the hyper-parameter $K$,  demonstrating that the fusing net can improve the accuracy and robustness of the model.

\begin{table}[th]
  \caption{The performance of the vanilla $k$NN and the fusing net.}
  \vspace{-6pt}
  \label{tab:ablation2}
  \centering
\begin{tabular}{c|cccc}
\toprule
 \multirow{2}{*}{$k$/K-size} & \multicolumn{2}{c}{U\_MSE↓} & \multicolumn{2}{c}{U\_KTAU↑} \\
 & Vanilla & Fusing & Vanilla & Fusing \\
 \midrule
5 & 0.216 & 0.197 & 0.695 & 0.704 \\
10 & 0.203 & 0.199 & 0.703 & 0.704 \\
15 & 0.198 & 0.197 & 0.707 & 0.705 \\
30 & 0.197 & 0.194 & 0.708 & 0.708 \\
60 & 0.197 & 0.195 & 0.708 & 0.708 \\
\midrule
mean & 0.202 & \textbf{0.196} & 0.704 & \textbf{0.706} \\
var & 6.57E-05 & \textbf{3.80E-06} & 3.07E-05 & \textbf{4.20E-06} \\
\toprule
\end{tabular}
\end{table}

\vspace{-16pt}
\section{Conclusions}
In this paper, we propose RAMP, a novel retrieval-augmented MOS prediction method, to enhance the neural decoder for alleviating the data scarcity issue in the SSL-based frameworks. We also design a fusing network to dynamically adjust the retrieval scope and the fusion weights based on the predictive confidence. The experimental results show that the proposed models perform well in both in-domain and cross-domain settings. 

\vspace{-8pt}
\section{Acknowledgements}
This work has been supported by the National Key R\&D Program of China through grant 2022ZD0116307 and NSF China (Grant No.62271270).

\bibliographystyle{IEEEtran}
\bibliography{mybib}

\end{document}